\documentstyle[twoside,fleqn,espcrc2]{article}

\newcommand{\beq}{\begin{equation}}
\newcommand{\eeq}{\end{equation}}
\newcommand{\bea}{\begin{eqnarray}}
\newcommand{\eea}{\end{eqnarray}}
\newcommand{\ep}{\epsilon}
\newcommand{\si}{\sigma}
\newcommand{\half}{{1\over 2}\;}
\newcommand{\ve}{\varepsilon}
\newcommand{\nn}{\nonumber}
\newcommand{\ie}{\hbox{\it i.e. }}

\newcommand{\AmS}{{\protect\the\textfont2
  A\kern-.1667em\lower.5ex\hbox{M}\kern-.125emS}}

\title{Correlation functions for the 2D random bonds Potts Models.}

\author{Vladimir Dotsenko\address{LPTHE\\
Universit\'e Pierre et
Marie Curie, PARIS VI - Universit\'e Denis Diderot, PARIS VII\\
Boite 126, Tour 16, 1$^{\it er}$ \'etage, 4 place Jussieu\\
F-75252 Paris CEDEX 05, FRANCE}
     \thanks{Also at the Landau Institute for
Theoretical Physics, Moscow}, Marco Picco$^a$ and Pierre
Pujol$^a$}

\begin{document}

\begin{abstract}
We study the spin-spin and energy-energy correlation functions for
the 2D Ising and 3-states Potts model with random bonds at the critical
point. The procedure employed is the renormalisation group approach of the
perturbation series around the conformal field theories representing the
pure models.  For the Ising model, we obtain a crossover in the amplitude
for the correlation functions which doesn't change the critical
exponent. For the $3$-state Potts model, we found a shift in the critical
exponent produced by randomness. A comparison with numerical data is
discussed briefly.
\end{abstract}

\maketitle

\section{INTRODUCTION}

For models with random bonds, the main problem is to determine if the
randomness leaves unchanged the critical properties of the pure system or
if the singularities of the thermodynamical functions are modified.
First results, which suggest an intermediate situation, have been obtained
by Harris and Lubensky \cite{harrlub}, Grinstein and Luther \cite{grins}
and Khmelnitskii \cite{khmel} using the standard $\phi^4$ theory. Other
cases, like long-range correlated quenched defects have also been
considered, (see for example \cite{weinrib,korzh}.) A first step in the
understanding of the relevance of randomness was given by the Harris
criterion \cite{harris}: the randomness is relevant (irrelevant) if the
specific heat exponent of the pure model is positive (negative). Two
dimensional systems are particularly interesting because of the rich
structure of conformal invariance in this dimension. Assuming that a random
model has a critical point with second order phase transition, the main
interesting problem is to determine which conformal field theory represents
this model at the infrared fixed point. Let us also mention that an exact
result has been obtained by McCoy and Wu \cite{mccoy} who considered a
two-dimensional Ising model where only vertical bonds on a square lattice
were allowed to acquire the same random value. They found that the
logarithmic singularity of the specific heat disappeared completely.

The models that we will study in this paper are the two dimensional Ising
and Potts models with random bonds. For the case of Ising model, Harris
criterion doesn't provide a qualitative answer of the relevance of the
randomness (the specific heat exponent for the $2-D$ Ising model is
$0$). First results for this model were obtained by Dotsenko and Dotsenko
\cite{dots1} who
showed that near the critical point, this model can be represented by an
$n=0$ Gross-Neveu model \cite{GN}. With this technique, they found that the
specific heat singularity get smoothed as $ln(ln({1\over|t|}))$ where $|t|$
is the reduced temperature. Calculation of spin-spin correlation function
by this technique which involves non-local fermionic representation of
$<\sigma \sigma>$, was later questioned by Shalaev and Shankar
\cite{shalaev,shankar} who gave arguments that the asymptotic behavior of
this correlation function is unchanged by the randomness (see also
\cite{ludwig2}.) Let's also mention some other approaches. In
\cite{ziegler}, the situation was questioned by Ziegler who claims that
non-perturbative effects introduce an intermediate phase around the
critical point of the pure model. In an other approach, Mussardo and
Simonetti studied the exact formulation of the random bond Ising model in
term of $S$-matrix \cite{mussardo}.
On the
other hand, numerical simulations of the Ising model \cite{talapov,adsw}
seem to confirm the theoretical predictions of the specific heat and
spin-spin correlation function asymptotic behavior.

For the Potts model,
the situation is more clear. Hear, the Harris criterion makes some precise
predictions: randomness is relevant and changes the
critical behavior.  Using conformal field theory techniques, Ludwig
\cite{ludwig1} also perturbatively computed a shift in the critical
exponent of the energy operator in the case of the random Potts model (this
critical exponent for the Ising model is also unchanged).

The paper is organized as
follows. In section $2$ we introduce the model. In section $3$, we briefly
present the methods for computing correlation functions, referring to
\cite{us} for more details. Results are presented in section $4$ while in
section $5$ a discussion of the results is presented and a comparison with
numerical data obtained recently.

\section{THE MODEL}

The partition function for a q-state Potts model with
a fixed configuration of disorder ${J_{ij}}$ is given by
\beq
Z(J_{ij})=\sum_{\{\sigma_i\}} e^{-\beta \sum_{<i,j>}
\delta_{\sigma_i,\sigma_j} J_{i,j}}
\eeq
Here, $\sigma_i$ corresponds to the value of the spin at the location $i$
on the lattice and takes the values
\beq
\sigma_i=0,\cdots,q-1
\eeq
and $J_{ij}$ corresponds to the coupling constants between neighboring
spins. These coupling constant can be separated in two parts
\beq
J_{ij}=J_0+\Delta_{ij}
\eeq
$J_0$ is the coupling constant of the pure model (without
disorder) while $\Delta_{ij}$ is the random part of the couplings.
This partition function can also be written under the form
\beq
Z(J_{ij})=\sum_{\{\sigma_i\}} e^{-S_0 -\beta \sum_{<i,j>}
\delta_{\sigma_i,\sigma_j} \Delta_{i,j}}
\eeq
where we have explicitly separate the non random part (to which corresponds
a action associated to a conformal field theory at the critical
temperature) and a random part. Thus, using the fact that the
$\delta_{\sigma_i,\sigma_j}$
corresponds, in the continuum limit of the pure Potts model, to the energy
field $\ep(z)$, we can also express the partition function like
\beq
Z(m)=Tr_{\{\sigma_i\}} e^{-S_0 -\int m(z)\ep(z) d^2z }
\eeq
and $m(z)$ is the continuum limit of $\Delta_{i,j}$. Under this form we
recognize a conformal field theory with a perturbation term
$\int d^2z\;  m(z)\ep(z)$.

The next step consists in performing the average over the disorder \ie over
$m(z)$. In fact, we need to compute the average over the free energy, \ie
over $\ln{Z(m)}$. This is done easily by using the replica method. Taking
the partition function
of $n$ identical copies of the system and analytically continuing to the
limit $n\to0$ gives the quenched free energy
$$
-\beta \overline{F} =
\overline{\ln(Z)} = \displaystyle\lim_{n\to0}{\overline{{Z^n - 1 \over n}}}
$$
where:
\begin{equation}
Z^n = \prod_{a=1}^n  Tr_{\{a,\sigma_i\}}
e^{-\displaystyle\sum_{a=1}^n S_{0,a} - \int m(z)\displaystyle\sum_{a=1}^n
\varepsilon_a(z) d^2z}
\end{equation}
the average of $ Z^n$ is made with a Gaussian distribution for $m(z)$:
$$
\overline{Z^n} = \int \prod_z dm(z) Z^n e^{-{1\over2g_0}(m(z)-m_0)^2}
$$
which gives:
\begin{equation}
\label{z}
\overline{Z^n} = \prod_{a=1}^n Tr_{a,s_i} e^{-\overline{S}}
\eeq
and
\bea
\overline{S}&=& \displaystyle\sum_{a=1}^n
S_{0,a} + g_0 \int \displaystyle\sum_{a,b=1}^n
\varepsilon_a(z)\varepsilon_b(z) d^2z \nn \\
&-& m_0 \int \displaystyle\sum_{a=1}^n
\varepsilon_a(z) d^2z
\eea
The terms in
$\int \displaystyle\sum_{a,b=1}^n \varepsilon_a(z)\varepsilon_b(z) d^2z$
containing the same replica label produce irrelevant operators and can be
omitted. Then $\overline{S}$ is reduced to
\bea
\overline{S}&=& \displaystyle\sum_{a=1}^n
S_{0,a} + g_0 \int \displaystyle\sum_{a\not=b}^n
\varepsilon_a(z)\varepsilon_b(z) d^2z \nn \\
&-& m_0 \int \displaystyle\sum_{a=1}^n
\varepsilon_a(z) d^2z
\eea
In the limit
$ m_0\rightarrow0$, this model corresponds to a conformal field theory
perturbed by a term quadratic in the $\varepsilon$
operator. Then the ``evolution'' of the coupling constants $g_0$ and $m_0$
under a renormalisation group (R.G.) transformation can be analyzed as well
as the behavior of the correlation functions.  In the calculation of
correlation
functions $<O(0)O(R)>$, where $O$ is some local operator, we will proceed
perturbatively:
\bea
<O(0)O(R)> &=& <O(0)O(R)>_0 \\
+<S_IO(0)O(R)>_0
&+&{1\over2}<S_I^2O(0)O(R)>_0+... \nn
\eea
where $<>_0$ means the expectation value taken with respect to $S_0$ and
\begin{equation}
S_I =
\int H_I(z) d^2z =  g_0 \int \displaystyle\sum_{a\not=b}
\varepsilon_a(z)\varepsilon_b(z) d^2z
\end{equation}
The operator $O$ is then renormalised as
$$
O\rightarrow O ( 1 + A_1g_0 + A_2g_0^2 + A_3 g_0^3 + \cdots ) \equiv
Z_0 O
$$
The integrals of correlation functions involved in the calculation can be
performed by analytic continuation with the Coulomb-gas representation of a
conformal field theory \cite{dots2} where the central charge is $c = {1
\over 2} +
\epsilon'$. The $\epsilon'$ term corresponds to a short distance regulator
for the integrals. In addition, we also used an infrared (I.R.) cut-off
$r$. The
result is then expressed as an $\epsilon'$ series with coefficients
depending on $r$. The limit $\epsilon' \rightarrow 0$ corresponds to
the pure Ising
model at the critical point while the Potts model is obtained for some
finite value of
$\epsilon'$. We recall here some notations of the Coulomb-gas representation
for the vertex operators \cite{dots2}. The central charge $c$ will be
characterized in the following by the parameter $\alpha^2_+ = {2p\over2p-1}
={4\over 3} + \ep $ with
\bea
\label{c}
c=1-24 \alpha_0^2 &;&  \alpha_\pm = \alpha_0 \pm
\sqrt{\alpha_0^2 +1} \\
\alpha_+ \alpha_- &=& -1 \nonumber
\eea
Note that for the pure 2D Ising model
 $\alpha_+^2 ={4\over 3} $ and $c = {1\over 2}$
while  for the $3$-state Potts model  $\alpha_+^2 =
{6\over 5}$, $c={4\over 5}$ and $\epsilon = -{2\over 15}$.
 The vertex operators are defined by
\begin{equation}
V_{nm}(x) = e^{i\alpha_{nm} \phi(x)}
\end{equation}
where $\phi(x)$ is a free scalar field and where the $\alpha_{nm}$ are
given by
\begin{equation}
\alpha_{nm}=\half (1-n) \alpha_- + \half (1-m) \alpha_+
\end{equation}
The conformal dimension of an operator $V_{nm}(x)$ is
$\Delta_{nm} =
-\alpha_{\overline{nm}} \alpha_{nm}$ with
\begin{equation}
\label{al}
\alpha_{\overline{nm}} = 2\alpha_0 -\alpha_{nm} =\half (1+n) \alpha_- +
\half (1+m) \alpha_+
\end{equation}
The spin field $\sigma$ can be represented by the vertex operator
$V_{p,p-1}$ whereas $V_{1,2}$ corresponds to the energy operator
$\varepsilon$. In the same way, we associate $ e^{i\alpha_+ \phi(x)}$ to
the screening charge operator $ V_+(x)$.
Note that in the Ising case the $\sigma$ operator can also be represented
by  the $V_{21}$ operator (since both operators coincide in the limit
$\epsilon \rightarrow 0$). So, we can represent our spin operator by
$V_{k,k-1}$ where $k = {2 + 3 \lambda  \epsilon \over 1 + 3
\epsilon}$. We have $\lambda = 2$ for $V_{21}$ and  $\lambda = {1\over2}$
for $V_{p,p-1}$.
\section{Renormalisation Group Equations}
In this section, we will deal with the computation of correlation functions
of operators $\varepsilon$ and $\sigma$. To compute them, one needs to
determine the effect of the random
coupling on the operators $\varepsilon$ and $\sigma$ and compute the
renormalised operators $\varepsilon '$ and $\sigma '$. This means that we
want to compute the functions $Z_\varepsilon$ and $Z_\sigma$ such that
\begin{equation}
\varepsilon ' = Z_\varepsilon \varepsilon \quad \mbox{and} \quad
\sigma ' = Z_\sigma \sigma
\end{equation}
A convenient way to define $Z_\varepsilon$ and $Z_\sigma$ is to
consider the more general action
\bea
\displaystyle\sum_{a=1}^n S_{0,a}
 -g_0 \int \displaystyle\sum_{a,b=1}^n \varepsilon_a(z)\varepsilon_b(z)
d^2z \nn \\
+ m_0 \int \displaystyle\sum_{a=1}^n \varepsilon_a(z) d^2z - h_0 \int
\displaystyle\sum_{a=1}^n \sigma_a(z) d^2z
\eea
This merely corresponds to the action used in (\ref{z}) with an additional
coupling of the $\sigma$ field. Then, with the help of the operator algebra
(O.A.) coming from contractions between $\ve$ and $\si$ operators, we will
compute the effect of the $g_0
\int \displaystyle\sum_{a,b=1}^n \varepsilon_a(z)\varepsilon_b(z) d^2z $
term (that we will denote $g_0(\varepsilon\varepsilon)$ in the following for
simplicity) on the coupling terms $m_0 \int
\displaystyle\sum_{a=1}^n \varepsilon_a(z) d^2z$ and $h_0 \int
\displaystyle\sum_{a=1}^n \sigma_a(z) d^2z$. More precisely, we will
compute
\bea
\sum_i \left( (g_0(\varepsilon\varepsilon))^i m_0 \int
\displaystyle\sum_{a=1}^n \varepsilon_a(z) d^2z \right)\nn\\
 \simeq m \int
\displaystyle\sum_{a=1}^n \varepsilon_a (z) d^2z
\eea
and
\bea
\sum_i \left( (g_0(\varepsilon\varepsilon))^i h_0 \int
\displaystyle\sum_{a=1}^n \sigma_a(z) d^2z \right) \nn\\
\simeq h \int
\displaystyle\sum_{a=1}^n \sigma_a (z) d^2z
\eea
$m$ and $h$ being the renormalised coupling constants. Obviously, this
computation will be perturbatively made only up
to some finite power in $g_0$. In fact, the first step of the computation
will be to determine the renormalised $g$
constant on which $Z_\varepsilon$ and $Z_\sigma$ depend.
\subsection{Renormalisation of the coupling constant $g$.}
The renormalisation of the coupling constant $g$ will be determined
directly by a perturbative computation. $g$ is also
given by the O.A. producing
\bea
g_0(\varepsilon\varepsilon)
&+&{1\over 2} \left(g_0(\varepsilon\varepsilon)\right)^2
+ {1\over 6} \left(g_0(\varepsilon\varepsilon)\right)^3 + \cdots \nn\\
&=&  g \int \displaystyle\sum_{a,b=1}^n
\varepsilon_a(z)\varepsilon_b(z) d^2z \nonumber
\eea
with $g=g_0 + A_2 g_0^2 + A_3 g_0^3 + \cdots $ where $A_2$ comes from
\bea
{1\over 2} \int \displaystyle\sum_{a,b=1}^n
\varepsilon_a(z)\varepsilon_b(z) d^2z \int \displaystyle\sum_{c,d=1}^n
\varepsilon_c(z)\varepsilon_d(z) d^2z \nn\\
= A_2 \int
\displaystyle\sum_{a,b=1}^n \varepsilon_a(z)\varepsilon_b(z) d^2z + \cdots
\eea
and $A_3$ from
\bea
{1\over 6} && \!\!\!\!\!\!\!\!\!\! \left( \int  \displaystyle\sum_{a,b=1}^n
\varepsilon_a(z)\varepsilon_b(z) d^2z \right)^3 \nonumber \\
&=& A_3 \int \displaystyle\sum_{a,b=1}^n
\varepsilon_a(z)\varepsilon_b(z) d^2z + \cdots
\eea
The computation of $A_2$ is made by contracting two $\varepsilon$
operators. We obtain (with $b=c\not= a,d$)
\bea
&{1\over 2}& \int \displaystyle\sum_{a,b=1}^n
\varepsilon_a(x)\varepsilon_b(x) d^2x \int \displaystyle\sum_{c,d=1}^n
\varepsilon_c(y)\varepsilon_d(y) d^2y \nn\\
&=&2(n-2)\int_{|x-y|<r}<\varepsilon(x)\varepsilon(y)>_0 d^2y \times\nn\\
& & \qquad \qquad \times \int \sum_{a,d=1}^n
\varepsilon_a(x)\varepsilon_d(x) d^2x\\
&=&4\pi(n-2)\int_{y<r}{dy\over y^{1+3\epsilon}} \int \sum_{a,d=1}^n
\varepsilon_a(x)\varepsilon_d(x) d^2x\nn\\
&=&-4\pi(n-2){r^{-3\epsilon}\over 3\epsilon}\int \sum_{a,d=1}^n
\varepsilon_a(x)\varepsilon_d(x) d^2x\nn
\eea
Thus $A_2=-4\pi(n-2){r^{-3\epsilon}\over 3\epsilon}$. Computations at
higher order will become very complicated. In the following, we will just
present the results, referring to \cite{us} for all the technical details.
For the computation of $g(r)$, calculations were made up to the third order
with the following result~:
\bea
g(r) &=& r^{-3\epsilon}( g_0 - g_0^2  4\pi (n-2){r^{-3\epsilon}\over
3\epsilon} \\
&+& g_0^3 8 \pi^2 (n-2) {r^{-6\epsilon}\over
3\epsilon} \left(1+ {2 (n-2)\over 3\epsilon}\right))\nn
\eea
Note that we multiply the result by $r^{-3\epsilon}$ in order to obtain
a dimensionless coupling constant $g(r)$.

{}From there, we can compute directly the $\beta$-function:
\bea
\beta(g) &=& {dg\over dln(r)} = -3\epsilon g(r) + 4 \pi (n-2) g^2(r) \nn\\
&-&16 \pi^2 (n-2) g^3(r) + O(g^4(r))
\eea
Finally, taking the limit $n \rightarrow 0$, we obtain for the
$\beta$-function up to the third order~:
\begin{equation}
\label{tobf}
\beta(g) =  -3\epsilon g - 8 \pi g^2 + 32 \pi^2 g^3
\end{equation}
{}From this $\beta$-function, we determine immediately the effect of the
disorder on the model. For the Ising model ($\epsilon=0$), the infrared
fixed point is $g=0$, while for the $3$-state Potts model ($\epsilon<0$), a
new infrared fixed point is reached with $g_c = -{3\epsilon \over 8
\pi} + {9\epsilon^2 \over 16\pi} + O(\epsilon^3) $.

\subsection{Renormalisation of $\sigma$ and $\varepsilon$}

In order to be able to compute the correlation functions of $\sigma$ and
$\varepsilon$, the second step is to determine the effect of the
renormalisation on these operators. One needs to compute the
multiplicative functions $Z_\sigma$ and $Z_\varepsilon$. This will be made
by computing the renormalised coupling terms $m\int
\displaystyle\sum_{a=1}^n \varepsilon_a (z) d^2z = (m_0 Z_\varepsilon )\int
\displaystyle\sum_{a=1}^n  \varepsilon_a(z) d^2z $ and
$h\int \displaystyle\sum_{a=1}^n \sigma_a (z) d^2z = (h_0
Z_\sigma)\int \displaystyle\sum_{a=1}^n \sigma_a(z) d^2z $.
A direct computation of both $m$ and $h$ will provide
us with the functions $Z_\sigma$ and $Z_\varepsilon$. As for the
computation of $g=Z_g g_0$, we will compute in perturbation~:
\bea
&m_0&\int\sum_{a=1}^n\ve_a(z) d^2z  + (g_0(\varepsilon
\varepsilon) m_0 \int \sum_{a=1}^n\ve_a(z) d^2z \nn\\
&+&
{1\over2}(g_0(\varepsilon \varepsilon))^2 m_0
\int\sum_{a=1}^n\varepsilon_a(z) d^2z  + \cdots \nn\\
&=& m\int\sum_{a=1}^n\ve_a(z) d^2z
\eea
and $m=m_0(1+B_1 g_0 + B_2 g_0^2 + \cdots)$ with $B_1$ defined by
\bea
\int \sum_{a,b=1}^n \ve_a(z)\ve_b(z) d^2z \int\sum_{a=1}^n\ve_a(z) d^2z
\nn\\
=
B_1 \int\sum_{a=1}^n\ve_a(z) d^2z
\eea
and $B_2$ by
\bea
\left(\int \sum_{a,b=1}^n \ve_a(z)\ve_b(z)\right)^2
\int\sum_{a=1}^n\varepsilon_a(z) d^2z \nn\\
= B_2 \int\sum_{a=1}^n\ve_a(z) d^2z
\eea
The details of the computation are presented in \cite{us}, with the
result
\bea
r^{-1+{3\over2} \epsilon}  m(r) = m_0 \Bigl( 1-4\pi
(n-1)g_0 {r^{-3\epsilon}\over
3\epsilon} \nn\\
+ 4 \pi^2 (n-1)g_0^2{r^{-6\epsilon}\over
3\epsilon} (1+{4n-6\over 3\epsilon} )\Bigr)
\eea
Here again, we multiply $m(r)$ by $r^{-1+{3\over2} \epsilon}$ in order to
obtain a dimensionless coupling constant. The R.G. equation for $Z_\ve$ is
thus given by
\begin{equation}
\label{eqzve}
{d ln(Z_\ve(r)) \over d ln(r)} =   4\pi (n-1) g
 -   8 \pi^2 (n-1) g^2
\end{equation}
Similarly, for the coupling constant $h_0$, we compute up to the third
order~:
\bea
\left(\int \sum_{a,b=1}^n \ve_a(z)\ve_b(z)\right)^i
\int\sum_{a=1}^n\si_a(z) d^2z \nn\\
= C_i \int\sum_{a=1}^n\si_a(z) d^2z
\eea
We give here directly the result :
\bea
&&\hskip -0.5cm r^{-{15\over8}-a(\epsilon)}h(r) =h_0 ( 1+ (n-1)g_0^2 \pi^2
{r^{-6\epsilon}\over 2}\times \\
&& \qquad \qquad \qquad (1 + {4\over3}(2-\lambda)
{\Gamma^2(-{2\over3})\Gamma^2({1\over6})\over\Gamma^2(-{1\over3})
\Gamma^2(-{1\over6})}) \nn\\
&& -12(n-1)(n-2) g_0^3 \pi^3
\left(r^{-9\epsilon}\over9\epsilon\right)\times \nn\\
&& \qquad \qquad \qquad (1+{8\over9}(2-\lambda)
{\Gamma^2(-{2\over3})\Gamma^2({1\over6})\over\Gamma^2(-{1\over3})
\Gamma^2(-{1\over6})})) \nn
\eea
The multiplicative term $r^{-{15\over8}-a(\epsilon)}$ in front of $h(r)$ is
introduced in order to make this parameter dimensionless. Here,
$a(\epsilon)$ is a function of $\epsilon$ depending on which representation
of the spin field we are taking in the Coulomb gas picture (see section
2). Its explicit form will be irrelevant in the following. The
corresponding R.G. equation for $Z_\si$ will be given by
\bea
\label{eqzsi}
&&\!\!\!\!\!\!\!\!\!\!\!\!{dln(Z_\si(r))\over dln(r)} = \\
&&\!\!\!\!\!\!\!\!\!\!\!\! -3(n-1)g^2(r) \pi^2 \epsilon
\left[1 + {4\over3}(2-\lambda)
{\Gamma^2(-{2\over3})\Gamma^2({1\over6})\over
\Gamma^2(-{1\over3})\Gamma^2(-{1\over6})}\right] \nn \\
&& \qquad \qquad \qquad \qquad + 4 (n-1)(n-2) \pi^3 g^3(r)\nn
\eea
\section{Correlation Functions}
We now have all the ingredients needed in order to compute the correlation
functions. They will be calculated with the help of the R.G. equations, for
the theory with $m_0$, $h_0\rightarrow 0$. From the R.G. equations, we
have~:
\bea
&&<\varepsilon(0)\varepsilon(sR)>_{r,g(r)} =\\
&&{Z^2_\ve(sr,g(sr))\over Z^2_\ve(r,g(r))}s^{-2\Delta_{\varepsilon}}
<\varepsilon(0)\varepsilon(R)>_{r,g(sr)}\nn
\eea
This can be written as~:
\bea
\label{zeps}
&&<\varepsilon(0)\varepsilon(sR)>_{r,g(r)} =\\
&&e^{2\int\limits_{g_0}^{g(s)}{\gamma_\ve(g)\over \beta(g)}
dg}s^{-2\Delta_{\varepsilon}} <\varepsilon(0)\varepsilon(R)>_{r,g(sr)}\nn
\eea
where we used the notation~:
\beq
\label{defzve}
{dlnZ_\ve\over dlnr} = \gamma_\ve(g)
\eeq
and $g(s) = g(sr); g_0 = g(r)$. We assume now $r$ to be a lattice cut-off
scale. In a similar way for $<\sigma(0)\sigma(R)>$ the R.G. equation is~:
\bea
\label{zeps2}
&&<\sigma(0)\sigma(sR)>_{r,g(r)} =\\
&&e^{2\int\limits_{g_0}^{g(s)}{\gamma_\si(g)\over \beta(g)}
dg}s^{-2\Delta_{\sigma}} <\sigma(0)\sigma(R)>_{r,g(sr)}\nn
\eea
with
\beq
\label{defzsi}
{dlnZ_\si\over dlnr} = \gamma_\si(g)
\eeq
In equations (\ref{zeps})-(\ref{zeps2}), $R$ is an arbitrary scale
which can be fixed to one lattice spacing $r$ of a true statistical model.
The dependence of $<\sigma(0)\sigma(r)>_{r,g(s)}$ on $s$ will then be
negligible, assuming that there are not interactions on distances
smaller than $r$. Therefore, it reduces to a constant. Then, $s$ will
measure the number of lattice spacings between two spins in
$<\sigma(0)\sigma(sR)>$. In the following, we adopt the choice $r=1$.
\subsection{The Ising model}
The Ising model corresponds to the case $\epsilon \rightarrow 0$ and so the
$\beta$ function is~:
\beq
\label{beti}
\beta(g) =  - 8 \pi g^2 + 32 \pi^2 g^3
\eeq
Therefore, we can see that the I.R. fixed point is located at $g=0$. Also
we have, by eqs.(\ref{eqzve}), (\ref{eqzsi}) for $n=0, \epsilon=0$ and
definitions (\ref{defzve}), (\ref{defzsi}),
\bea
\gamma_\varepsilon (g)&=&-4\pi g + 8\pi^2 g^2 \\
\gamma_\sigma (g)&=&8\pi g^3
\eea
The integral for the $\varepsilon$ correlation
function, eq.(\ref{zeps}), gives~:
\bea
2\int\limits_{g_0}^{g(s)}{\gamma_\ve(g)\over \beta(g)} dg &=&
\int\limits_{g_0}^{g(s)} {1-2\pi g\over 1- 4\pi g}{dg\over g}\nn\\
&\approx&  \int\limits_{g_0}^{g(s)} (1+2 \pi g){dg\over g}\\
&=& 2\pi(g(s)-g_0)+ ln\left(g(s)\over g_0 \right)\nn
\eea
Now, we need to compute $g(s)$. The integration of equation
$\beta(g)={dg\over dln(r)}$ gives~:
$$
\int\limits_{g_0}^{g(s)} {dg\over -8\pi g^2 + 32 \pi^2 g^3} =
\int\limits_{r}^{sr} dlnr
$$
with the following solution up to the second order~:
\bea
&&\hskip -0.7cm g(s) =\\
&&\hskip -0.7cm {g_0\over 1+8\pi g_0 ln(s)}(1 + {4\pi g_0 ln(1+8\pi g_0
ln(s))\over
1+8\pi g_0 ln(s) })+ O(g_0^3)\nn
\eea
So, $<\varepsilon \varepsilon>$ correlation function is given by~:
\bea
&&\hskip -0.7cm <\varepsilon(0) \varepsilon(s)>_{g_0} \sim {g(s)\over g_0}
\left( 1 -
2\pi(g_0 - g(s))\right) s^{-2\Delta_{\varepsilon}} \nn\\
&&\hskip -0.7cm \sim {1\over 1+8\pi g_0 ln(s)}s^{-2\Delta_{\ve}}(1+
{4\pi g_0\over 1+8\pi g_0 ln(s)}\times\\
&&(ln(1+8\pi g_0 ln(s)) - 4\pi g_0 ln(s)))  + O(g_0^2)
\nn
\eea
For the $\sigma$ correlation function, the computation is similar. In fact,
in that case, we have $\gamma_\si(g)=8\pi^3 g^3$. Thus, keeping only the
first order of the $\beta$-function (\ie $\beta(g) =  - 8 \pi g^2$), we
obtain~:
\bea
&&2\int\limits_{g_0}^{g(s)}{\gamma_\si(g)\over \beta(g)} dg =
-2\pi^2 \int\limits_{g_0}^{g(s)}gdg\nn\\
&&
= -\pi^2 \left( g(s)^2 - g_0^2 \right) + O(g_0^3)
\eea
The $<\sigma \sigma>$ correlation function is then found to be given by~:
\bea
&&\hskip -0.9cm <\sigma(0) \sigma(s)> \sim \left( 1 + \pi^2  \left( g_0^2
- g(s)^2 \right) \right)  s^{-2\Delta_{\sigma}} \\
&& \hskip -0.9cm \sim
 \left( 1 + \pi^2 g_0^2  \left( 1 - {1\over (1+8\pi g_0 ln(s)
)^2 }\right)\right)
 s^{-2\Delta_{\sigma}}+O(g_0^3) \nn
\eea
The calculation of the $g^3$ term in the $\beta$ function and the $g^2$
term in the renormalisation of $\varepsilon$ was already done in
\cite{ludwig1}, extending the one loop result of \cite{dots1}. We recovered
these higher order corrections using a different technique, which allowed
us to calculate also the modified correlation function of the spin
operators.

\subsection{The Potts model}
We consider here the 3-state Potts model. With our conventions this case
corresponds to $\epsilon = -{2\over15}$. $\beta(g)$ is given in
eq.(\ref{tobf}), and
\bea
\gamma_\varepsilon (g) &=& -4\pi g + 8\pi^2 g^2 \\
\gamma_\sigma (g) &=& 3\pi ^2 \epsilon \left( 1 +{4\over 3}(2-\lambda)
{\Gamma^2(-{2\over 3}) \Gamma^2({1\over 6}) \over \Gamma^2(-{1\over 3})
\Gamma^2 (-{1\over 6})}\right) g^2 \nn\\&+& 8\pi^3 g^3
\eea
At long distances, the integrals in eqs.(\ref{zeps}),
(\ref{zeps2}) will be dominated by the
region $g \sim g_c$, with $g_c = -{3\epsilon \over 8
\pi} + {9\epsilon^2 \over 16\pi} + O(\epsilon^3) $. This is
different from the Ising model, because here
$\gamma_\ve(g_c)$ and $ \gamma_\si(g_c)$ have finite values for
$g=g_c$. Thus,
\beq
\int\limits_{g_0}^{g(s)}{\gamma_\ve(g)\over \beta(g)} dg \approx
\gamma_\ve(g_c) ln(s)
\eeq
and
\beq
\int\limits_{g_0}^{g(s)}{\gamma_\sigma(g)\over \beta(g)} dg \approx
\gamma_\ve(g_c) ln(s)
\eeq
The correlation functions can then be deduced directly~:
\beq
<\varepsilon(0) \varepsilon(s)> \sim
s^{-(2\Delta_{\varepsilon}-2\gamma_\ve(g_c)) }
\eeq
and
\beq
<\sigma(0) \sigma(s)> \sim
s^{-(2\Delta_{\sigma}-2\gamma_\si(g_c)) }
\eeq
So, we can see that a direct consequence of the new IR fixed point is a
modification of the critical exponents $\Delta_{\varepsilon}$ and
$\Delta_{\sigma}$. A straightforward computation will give these new
exponents~:
\bea
2\Delta'_{\varepsilon} &=& 2\Delta_{\varepsilon} - 2\gamma_\ve(g_c) \nn\\
&=&
2\Delta_{\varepsilon} + 8\pi g_c - 16 \pi^2 g_c^2\nn\\
 &=& 2\Delta_{\varepsilon} -3\epsilon
+{9\over4}\epsilon^2 + O(\epsilon^3)
\eea
and
\bea
2\Delta'_{\sigma} &=& 2\Delta_{\sigma} - 2\gamma_\si(g_c)  \\
&=& 2\Delta_{\sigma} - 6 \pi^2 g_c^2 \epsilon (1 \nn\\
&&+ {4\over3}(2-\lambda)
{\Gamma^2(-{2\over3})\Gamma^2({1\over6})\over\Gamma^2(-{1\over3})
\Gamma^2(-{1\over6})}) - 16 \pi^3 g_c^3 \nn\\
&=& 2\Delta_{\sigma} - {9\over8} (2-\lambda)
{\Gamma^2(-{2\over3})\Gamma^2({1\over6})\over\Gamma^2(-{1\over3})
\Gamma^2(-{1\over6})} \epsilon^3 + O(\epsilon^4)\nn
\eea
Here $\lambda = {1\over2}$ and the final result for the new critical
exponent is~:
\beq
\label{ndim}
2\Delta'_{\sigma} = 2\Delta_{\sigma} - {27\over16}
{\Gamma^2(-{2\over3})\Gamma^2({1\over6})\over\Gamma^2(-{1\over3})
\Gamma^2(-{1\over6})} \epsilon^3 + O(\epsilon^4)
\eeq
The value of the critical exponent for the energy operator was already
computed, up to the second order, by Ludwig \cite{ludwig1}. While for the
pure case $\Delta_\varepsilon=0.8$, here we obtained
$\Delta_\varepsilon'=1.02 + O(\epsilon^3)$. The difference between these
exponents is quite important and should be measured in a numerical
simulation of this model. See next section for a discussion on this point.
For the spin
operator, deviation of the critical exponent from the pure case appears
only at the third order, which we compute. This deviation is in fact very
small. While in case of 3-state Potts model without disorder
$2\Delta_{\sigma} = {4\over 15}$, we obtain the new  critical
exponent $2\Delta'_{\sigma} = {4\over 15} + 0,00264 = 0,26931$. Thus, the
deviation corresponds to a modification of $1\%$.
\section{Discussion}
In case of Ising model spin-spin function the calculation of up to third
order of the renormalisation group was needed to find the deviation from
the perfect model case, in the form of the cross-over in the
amplitude. This completes the observations of
\cite{shalaev,shankar,ludwig2}, based on absence of renormalisation of this
function in the first order, that asymptotically the spin-spin function has
the same exponent as in case of the perfect lattice model.

Recently the numerical simulations of the random Ising model has been
performed which measure directly  the deviation of $<\sigma \sigma> $ from
the pure Ising model at the critical point \cite{talapov}. These
measurements were made for disorder such that
$8\pi g_0 \approx 0.3$. Deviations predicted by our
computations are very small. They correspond to $0.1 \%$. The deviations
obtained in numerical simulations are around ten times larger, and they are
of opposite sign, \ie $ $ correspond to an extra decrease of the spin-spin
function
with distance $r$. In \cite{talapov}, it has been checked that this
decrease corresponds, within the accuracy of the measurements, to a factor
function of the ratio ${r/L}$, $F({r/L})$, $r$ being the distance between
the spins and $L$ is the lattice size. So they correspond to finite size
effects, being different for perfect and random models. We would suggest,
on the bases of our calculation of the $r$ dependence of the
spin-spin function on an infinite lattice, that numerical deviations will
continue to be plotted by the same curve $F({r/L})$, if one measures
$<\sigma \sigma>$ for different lattice sizes as it has been done in
\cite{talapov}, until the accuracy reaches the value of the $r$-deviation
which we calculated here. Only then the curves for different $L$ will
split.

Recently numerical simulations of the $3$-state Potts model with disorder
were performed by one of the authors \cite{usf}. In these simulations,
$\Delta_{\sigma'}$ was measured as well
as $\Delta_{\varepsilon'}$. For $\Delta_{\sigma'}$ , it is very difficult
to obtain any conclusion. The measured value $\Delta_{\sigma'}$ is very
close to $\Delta_{\sigma}$ and error bars do include the case
$\Delta_{\sigma'}=\Delta_{\sigma}$ up to a deviation of $1\%$, which is the
deviation predicted by analytical computations presented in this work. For
the energy-energy correlation function, the measured exponent is $
\Delta_{\varepsilon'}=1.065\pm 0.02$, to be compared with the prediction of
our computations: $\Delta_{\varepsilon'}=1.02$. This is quite a
satisfactory result if we remember that the result
$\Delta_{\varepsilon'}=1.02$ is a perturbative computation while the pure
model (around which we perturbed) has the value $\Delta_{\varepsilon}=0.8$

\end{document}